\documentclass{article}
\usepackage[letterpaper, margin=1in]{geometry}
\usepackage{float}
\usepackage[font=small,labelfont=bf]{caption}
\usepackage{amsmath}
\numberwithin{equation}{section}
\usepackage{amsmath,latexsym}
\usepackage{amssymb}
\usepackage{graphicx}
\usepackage{multicol}
\usepackage{subcaption}
\usepackage[utf8]{inputenc}
\pdfoutput=1
\usepackage{hyperref}
\hypersetup{
  pdfinfo={
    Title={Your Title Here},
    Author={Your Name Here},
    Subject={If you want to put something here, do so},
    Keywords={Add some keywords if you feel so inclined}
  }}
\usepackage{pdfpages}
\usepackage{microtype}

\title{Matrix Model: Emergence of a Quantum Number\\ in the Strong Coupling Regime}
\author{Castaly Fan, Larry Zamick\\
\vspace{1.4em}
\\
Department of Physics and Astronomy, Rutgers University, \\
 Piscataway, New Jersey 08854, USA}

\begin{document}

\maketitle

\begin{abstract}
    We continue here to study simple matrix models of quantum mechanical Hamiltonians. The eigenvalues and eigenfunctions were associated energy levels and wave functions. Whereas previously we considered the weak coupling limits of our models, we here address the more difficult strong coupling limits. We find that the wave functions fall into two classes and we can assign a quantum number to distinguish them. Implications for transition rates are also discussed.
\end{abstract}

\section{Introduction}
    In previous works we studied the properties of simple tridiagonal and pentadiagonal matrices \cite{1}\cite{2}\cite{3}\cite{4}\cite{5}. We here show the complete pentadiagonal matrix with 2 parameters, $v$ and $w$. We can reduce it to a tridiagonal case by simply setting $w$ to zero. Indeed in this work we will focus only on tridiagonal matrices.
    \begin{equation}
    H_{c} =
    \begin{pmatrix}
    0 & v & w & 0 & 0 & 0 & 0 & 0 & 0 & 0 & 0 \\
    v & E & v & w & 0 & 0 & 0 & 0 & 0 & 0 & 0 \\
    w & v & 2E & v & w & 0 & 0 & 0 & 0 & 0 & 0 \\
    0 & w & v & 3E & v & w & 0 & 0 & 0 & 0 & 0 \\
    0 & 0 & w & v & 4E & v & w & 0 & 0 & 0 & 0 \\
    0 & 0 & 0 & w & v & 5E & v & w & 0 & 0 & 0 \\
    0 & 0 & 0 & 0 & w & v & 6E & v & w & 0 & 0 \\
    0 & 0 & 0 & 0 & 0 & w & v & 7E & v & w & 0 \\
    0 & 0 & 0 & 0 & 0 & 0 & w & v & 8E & v & w \\
    0 & 0 & 0 & 0 & 0 & 0 & 0 & w & v & 9E & v \\
    0 & 0 & 0 & 0 & 0 & 0 & 0 & 0 & w & v & 10E \\
    \end{pmatrix}
    \label{1}
    \end{equation}

\begin{figure}[H]
        \centering
        \captionsetup{width=0.7\linewidth}
        \includegraphics[width=0.55\textwidth]{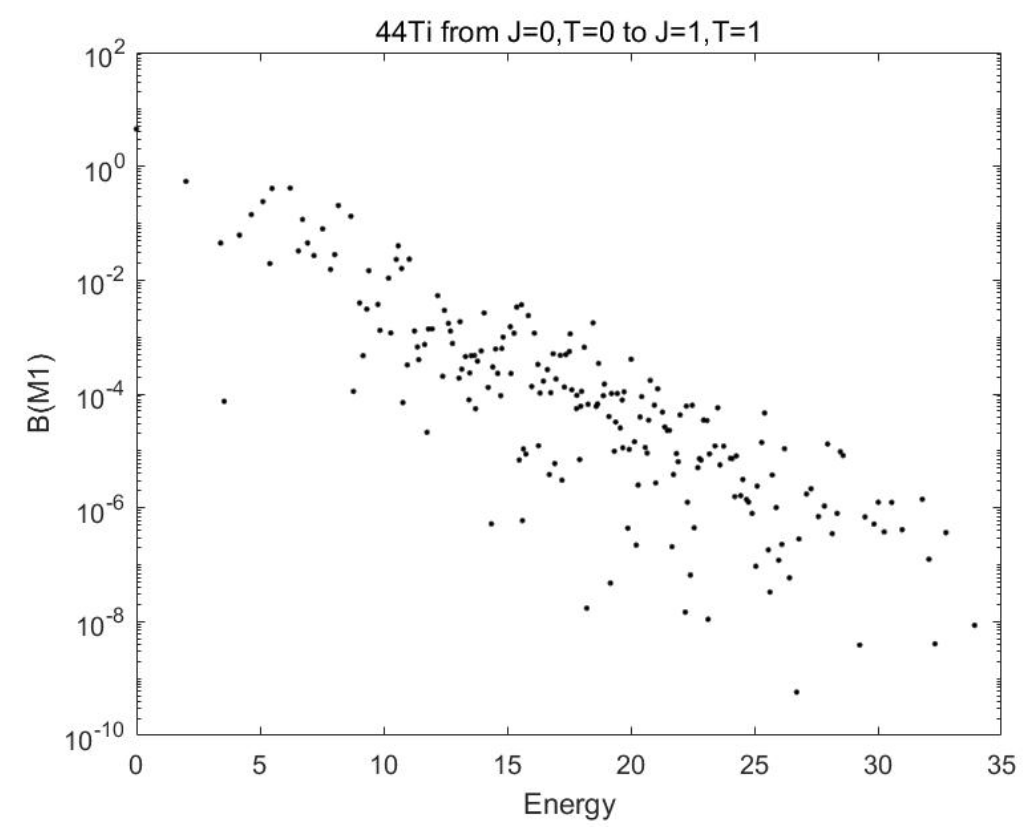}
        \caption{$^{44}$Ti B(M1)'s from Lowest $J=0$ $T=0$ to Lowest 205 $J=1$ $T=1$, Log Scale, without binning.}
        \label{fig1}
    \end{figure}

\begin{figure}[H]
        \centering
        \captionsetup{width=0.7\linewidth}
        \includegraphics[width=0.55\textwidth]{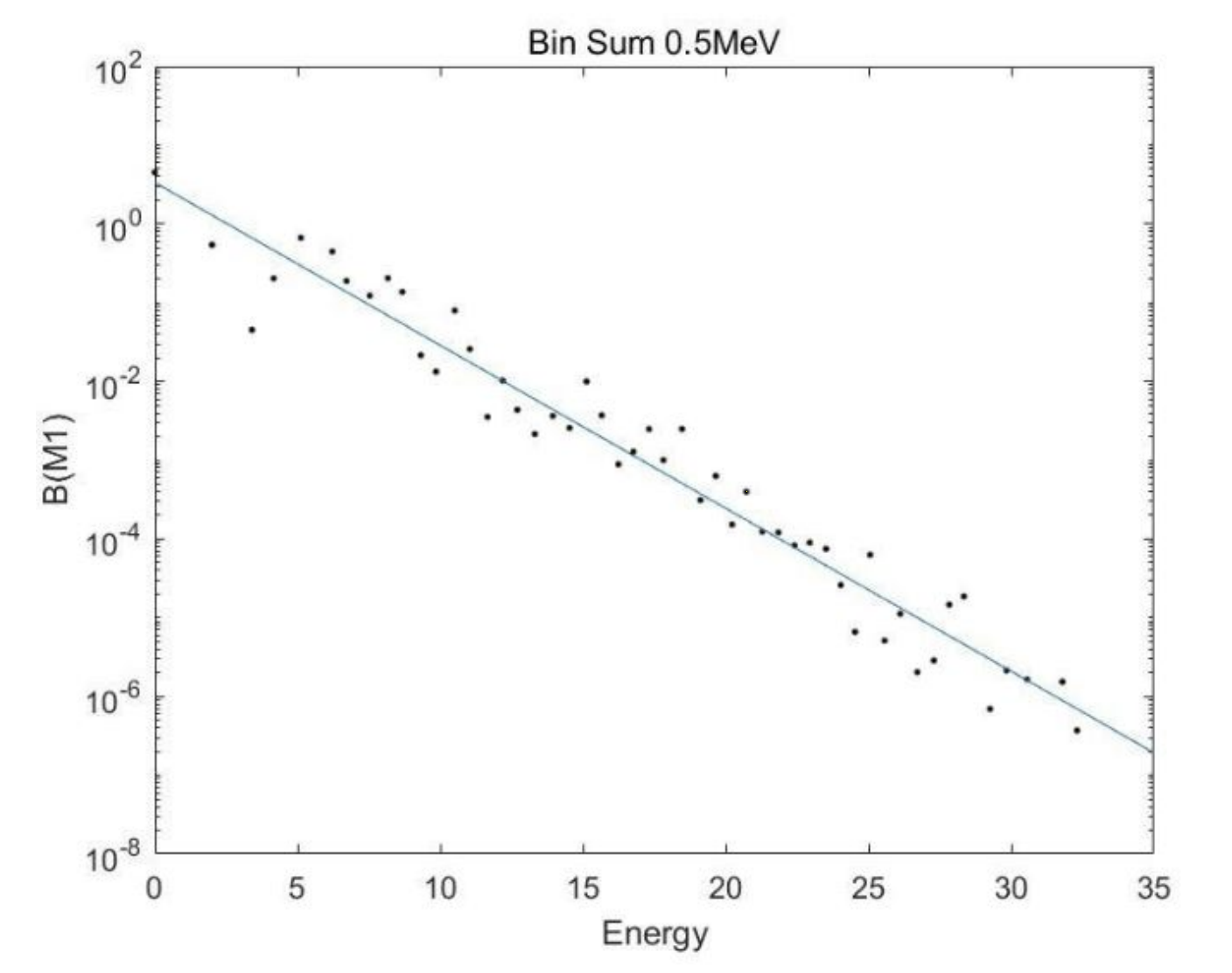}
        \caption{$^{44}$Ti B(M1)'s from Lowest $J=0$ $T=0$ to Lowest 205 $J=1$ $T=1$, Log Scale, with binning.}
        \label{fig2}
    \end{figure}

    Part of the motivation was the observation of an exponential decrease in calculated magnetic dipole strength \cite{6} in a realistic calculation with the NushellX program \cite{7} and the GXPF1A interaction \cite{8} We show what is mean by showing two figures which are based on the work of Kingan, Ma, and Zamick \cite{6}. In Figure \ref{fig1} we show the B(M1)'s from the lowest $J=0$ $T=0$ state in $^{44}$Ti to the lowest 205 $J=1^{+}$ $T=1$ states. There is much scatter in the results, but on a log plot (base $e$) we can crudely see an exponential decrease in strength with increasing excitation energy. In Figure \ref{fig2} we obtained the summed strength in 0.5 MeV bins. This shows much less scatter and a least fit straight line with a negative slope is drawn through the data. In ref. \cite{6} several other nuclei and several other $[J_{i}\; T_{i}]$ to $[J_{f}\; T_{f}]$ channels are considered and they all show exponential decreases in B(M1) strengths after binning.
    
    Somewhat motivated by the above binning results, we defined two different transition operators for use with the matrix (\ref{1}). To do this we first introduce the symbol $a(m,n)$. When matrix diagonalization is performed the outputs are eigenvalues which we associate with energy levels and eigenfunctions who we associate with wave functions. The amplitude of the $n$th component of the $m'$th eigenfunction is denoted as $a(m,n)$. Note the normalization and orthogonality conditions:
    \begin{gather*}
        \sum_{n} \:\lvert a(m,n) \rvert^{2} = 1\\
        \sum_{n} a(m,n)\:a(m',n) = \delta(m,m').
    \end{gather*}

\begin{figure}[H]
     \centering
     \begin{subfigure}[b]{0.45\textwidth}
         \centering
         \includegraphics[width=\textwidth]{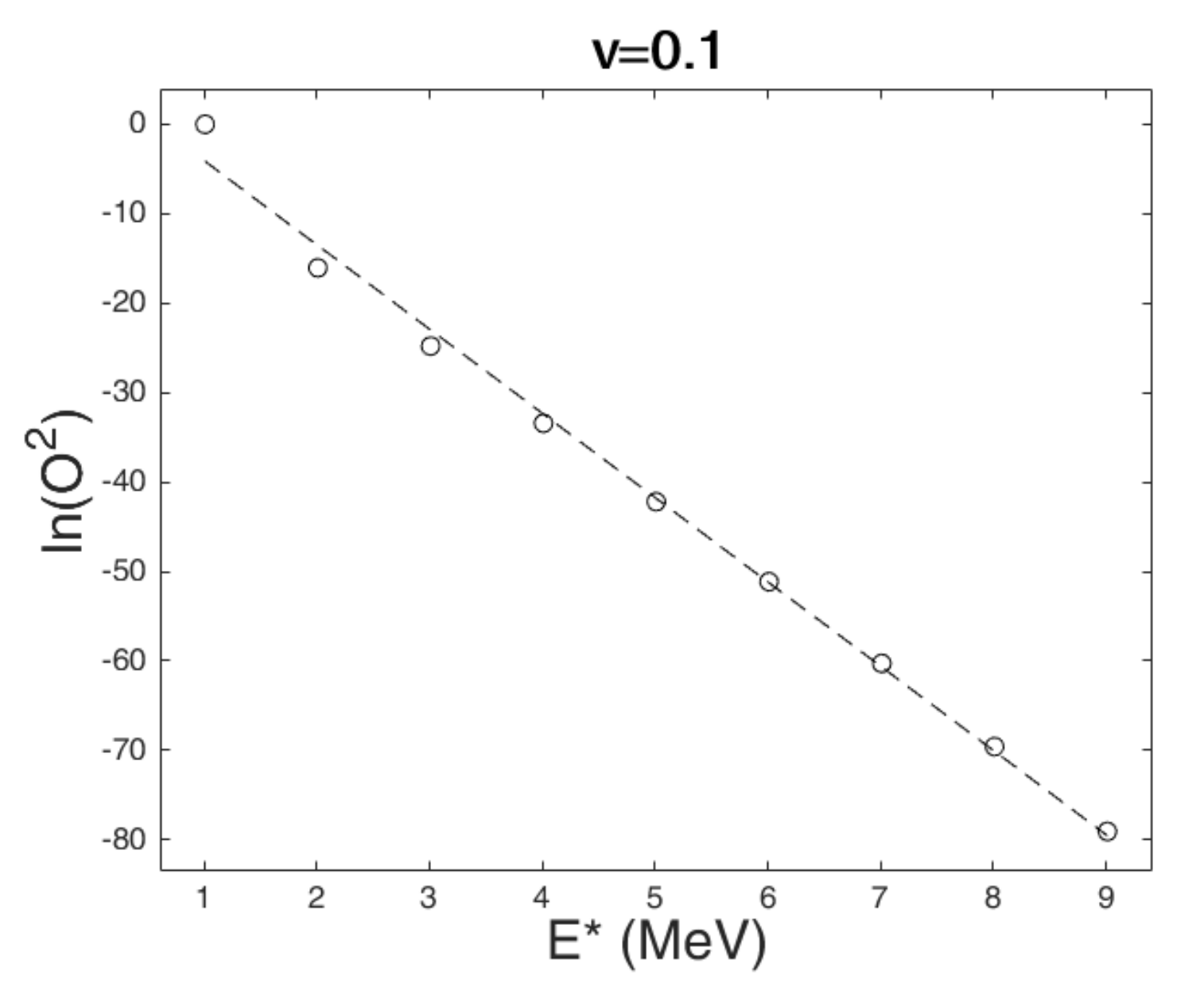}
     \end{subfigure}
     \quad
     \begin{subfigure}[b]{0.45\textwidth}
         \centering
         \includegraphics[width=\textwidth]{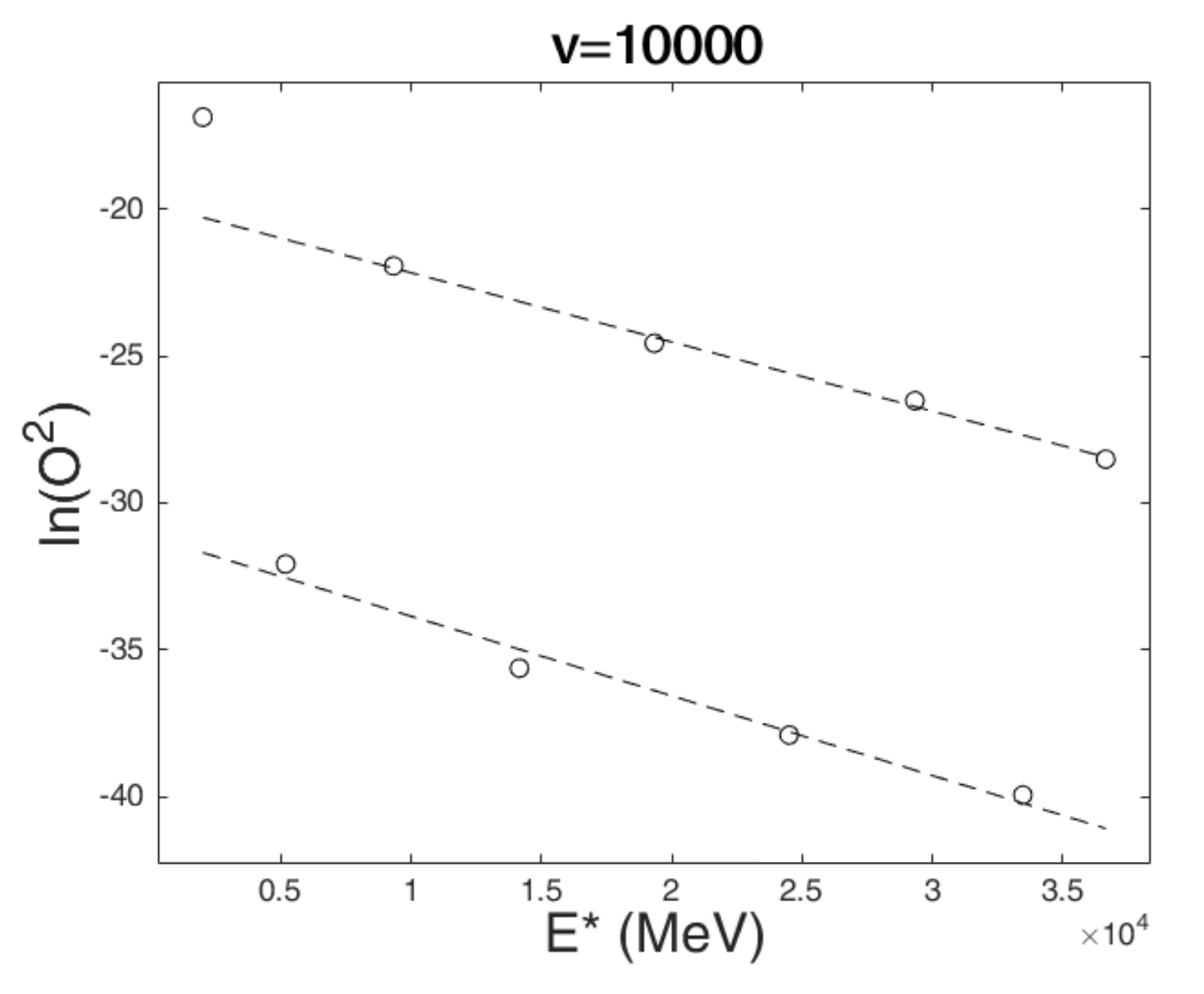}
     \end{subfigure}
     \caption{Log plot of transition rates from the lowest to all excited states state both in the weak coupling case ($v=0.1$) and the strong coupling case ($v = 10000$).}
    \label{fig3}
    \end{figure}

    We define two types of transition operators T1 and T2. For transitions from state $m$ to state $n$ the respective transition amplitudes are as follows:
    \begin{equation}
    \begin{split}
        \text{T1}: O(m,n) =  &a(m,0) a(n,1) +...+a(m,9) a(m,10)\\ 
                  &+ a(m,1) a(n,0) +...+a(m,10) a(n,9)\\
        \text{T2}: O(m,n) = &a(m,0) a(n,1) \sqrt{1} +...+ a(m,9) a(n,10)\sqrt{10}\\
                  &+ a(m,1) a(n,0) \sqrt{1} +...+ a(m,10) a(n,9) \sqrt{10}.
    \end{split}
    \end{equation}
    The transition rate from state $m$ to $n$ is then $O(m,n)^2$. As an example of we show in Figure \ref{fig3} a log plot of transition rates from the ground state to excited states both for small $v$ (0.1) and large $v$ (10,000) using the T1 transition operator. This is taken from ref. \cite{2} and shows for small $v$ the exponential decrease of transient strength with excitation energy. However as $v$ becomes larger, it becomes more evident that their is an even-odd effect, with transitions from ground to odd $n$ states being much stronger than to even $n$ states. We get two exponential decay lines.
    
\begin{figure}[H]
     \centering
     \begin{subfigure}[b]{0.45\textwidth}
         \centering
         \includegraphics[width=\textwidth]{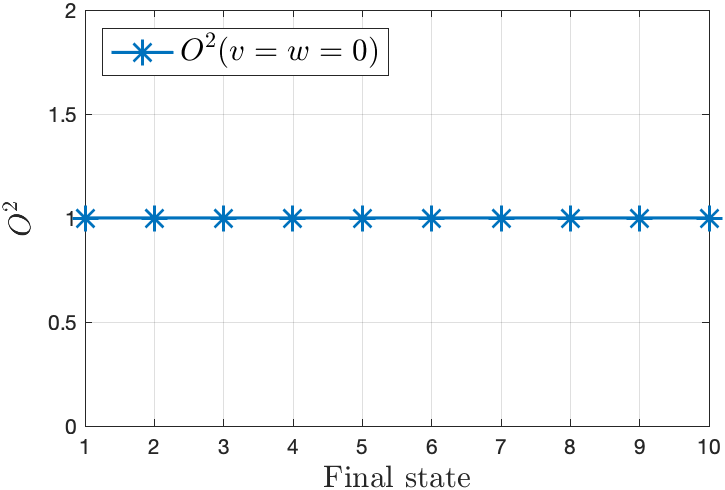}
     \end{subfigure}
     \quad
     \begin{subfigure}[b]{0.45\textwidth}
         \centering
         \includegraphics[width=\textwidth]{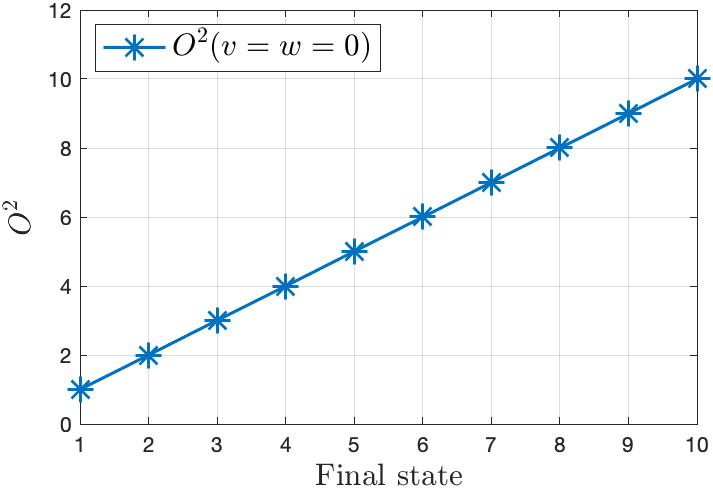}
     \end{subfigure}
     \caption{The transition strength $O^{2}$ for the flat case (left) and the harmonic case (right). Here the horizontal axis is with respect to the energy difference $E(n+1)-E(n)$.}
    \label{fig4}
    \end{figure}

    The difference between T1 and T2 is best shown for the case $v=0$, $w=0$ where we show figures of the $(m+1)$ to $m$ transitions rates. As seen in Figure \ref{fig4} (left) for T1, the curve is  horizontal (flat) with all $(m+1)$ to $m$ transition rates equal to one. For T2 which we call harmonic we have in Figure \ref{fig4} (right), we have a straight line such that the $(m+1)$ to $m$ rates is simply $(m+1)$.

    In this initial work of Kingan and Zamick \cite{1}, it was noted that the strength distribution, $O(0,n)^{2}$ vs. $n$ showed an exponential decrease with $n$, reminiscent of the above mentioned binning behavior of ref. \cite{6}. This lead to the thought that these simple matrix models were worth further study.

   In ref. \cite{4} besides excitation energy plots we also consider top-down transitions. That is to say, we start with the highest energy state and calculate the cascade of transitions until the ground states is reached. On a log plot we show the average transition strength as a function of the number of energy intervals that are crossed. This is motivated by experiments and theories by several groups \cite{9}\cite{10}\cite{11}\cite{12}\cite{13}. In previous works \cite{3}\cite{4}\cite{5} we studied the weak coupling limit of our model ($v$ much smaller than $E$). This is shown by comparing Figure \ref{fig2} and Figure \ref{fig3}. In the Figure \ref{fig2}'s log plot, we see close to a single line with a negative slope. This is for a small value, $v=0.1$. But for $v= 10,000$ we see close to two lines, one for even $n$ and one for odd $n$, which is an even-odd effect. In this work we will study this problem more in depth. That is to say we study the strong coupling limit of our model. At first thought this should be much more difficult to handle than weak coupling and to a large extent it is. However we shall show that there are some simple and striking systematics even here.

\section{The matrix model and symmetric patterns}
    As mentioned in the introduction our model consists of a Hamiltonian represented by the 11 by 11 matrix as Eq (\ref{1}).
    We know that different values of $v$ might lead to different values of the wave function. 
    
   We can reach the strong coupling limit by setting the diagonal terms of our matrix to zero. We will here limit ourselves to a tridiagonal matrix with only one parameter $v$, as shown in Eq (\ref{2}) : 
    \begin{equation}
    H_{a} =
    \begin{pmatrix}
    0 & v & 0 & 0 & 0 & 0 & 0 & 0 & 0 & 0 & 0 \\
    v & 0 & v & 0 & 0 & 0 & 0 & 0 & 0 & 0 & 0 \\
    0 & v & 0 & v & 0 & 0 & 0 & 0 & 0 & 0 & 0 \\
    0 & 0 & v & 0 & v & 0 & 0 & 0 & 0 & 0 & 0 \\
    0 & 0 & 0 & v & 0 & v & 0 & 0 & 0 & 0 & 0 \\
    0 & 0 & 0 & 0 & v & 0 & v & 0 & 0 & 0 & 0 \\
    0 & 0 & 0 & 0 & 0 & v & 0 & v & 0 & 0 & 0 \\
    0 & 0 & 0 & 0 & 0 & 0 & v & 0 & v & 0 & 0 \\
    0 & 0 & 0 & 0 & 0 & 0 & 0 & v & 0 & v & 0 \\
    0 & 0 & 0 & 0 & 0 & 0 & 0 & 0 & v & 0 & v \\
    0 & 0 & 0 & 0 & 0 & 0 & 0 & 0 & 0 & v & 0 \\
    \end{pmatrix}
    \label{2}
    \end{equation}
    
    In Table \ref{tab1} we show the wave functions for the Hamiltonian Eq (\ref{2}) i.e. for the strong coupling limit for positive $v$. Note that the diagonal terms are zero in (\ref{2}). The wave functions do not depend on what $v$ is. But the energy levels do and we show them for $v= 10^4$. 
    
    \begin{table}[H]
    \centering
    \captionsetup{width=0.85\linewidth}
    \caption{Wave functions $a(m,n)$ in the strong coupling limit (independent of positive $v$) and energy levels for $v=10^4$, $w=0$, given by solutions for Hamiltonian (\ref{2}).}
    \includegraphics[width=1\textwidth]{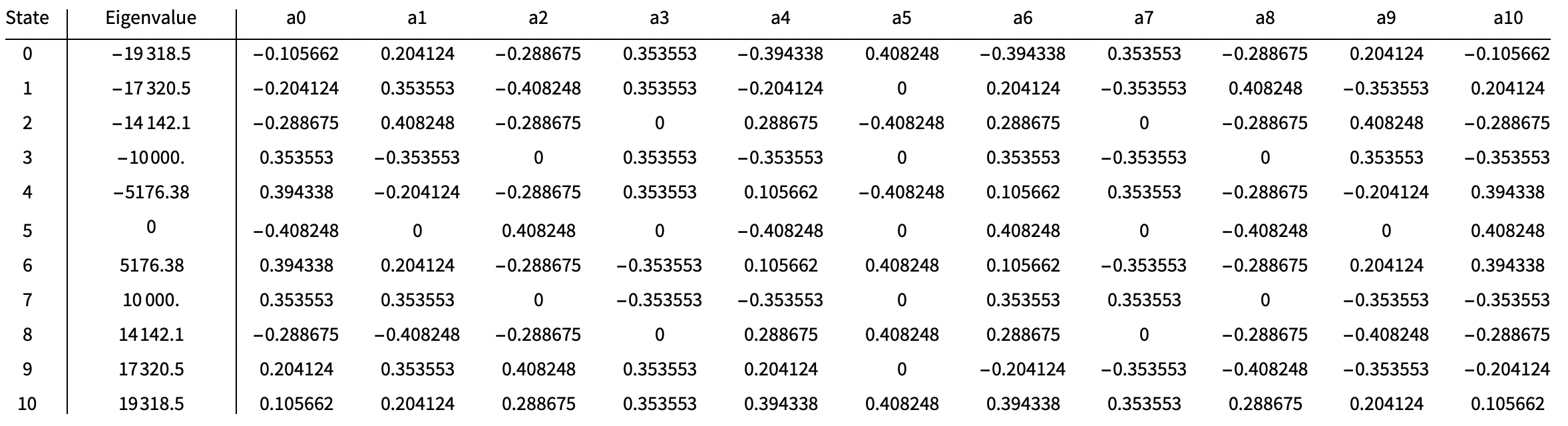}
    \label{tab1}
    \end{table}
    
    Next, in Table \ref{tab2} we get the corresponding wave functions for the full Hamiltonian (\ref{1}). Here the diagonal terms $nE$ are present with $E=1$, $v =10^4$ and $w=0$. We now have strong coupling but we are not at the strong coupling limit.

    \begin{table}[H]
    \centering
    \captionsetup{width=0.85\linewidth}
    \caption{Wave functions and energy levels for the Hamiltonian (\ref{1}) with $v=10^4$, $w=0$, $E=1$.}
    \includegraphics[width=1\textwidth]{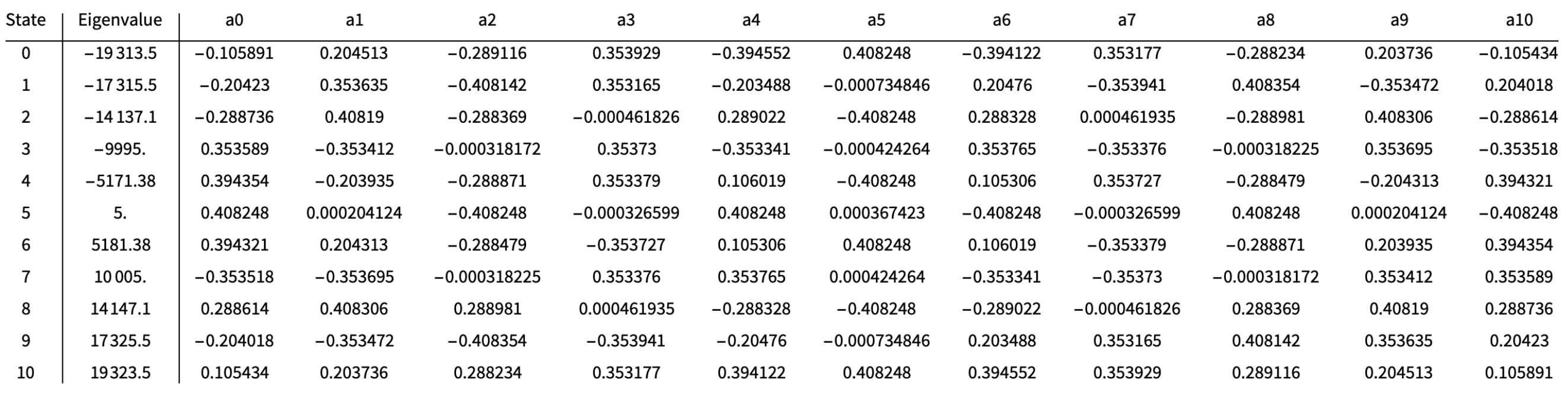}
    \label{tab2}
    \end{table}
    
    In Table \ref{tab2} we show the wave functions for $v=10^4$, and the diagonal terms $E_{n}= nE$ (with $E=1$). We no longer have the symmetry where e.g. the magnitude of $a_0$ is equal to the magnitude of $a_{10}$. The former is 0.105891 while the latter is 0.105434. However there still are some symmetries. The magnitudes of $a(0,n)$ are the same as the magnitudes $a(10,10-n)$  where $n$ is less than or equal to 5. For instance, $a(0,1)$ and $a(10,9)$ have the same magnitude of 0.204513. In other words, ignoring the phases the wave function components for $m=10$ are in reverse order as those for $m=0$; likewise $m=9$ and 2, 8 and 3, etc.

    We note the wave functions fall into 2 classes. Let $m$ represent the $m$'th wave function. For even  $m$  wave functions  we have  $a(m,1) = a(m,10)$; for odd $m$  wave functions  we have $a(m,1) = - a(m,10)$.  In more detail for even $m$, $a(m,n)= a(10-m,10-n)$; while for odd $m$, $a(m,n) = -a(10-m,10-n)$.  Here $n$ is less than or equal to 5.  
    
    What this shows is that in the strong coupling limit there is an interlacing of 2 types of wave functions. We can thus define an asymptotic quantum number 
    \begin{equation}
        s(m) = \frac{a(m,10)}{a(m,0)}.
    \label{2.2}
    \end{equation}
    There are 2 possible values: $+1$ for even $m$ and $-1$ for odd $m$.

\section{The quantum number under strong coupling}
    
    From Eq \ref{2} we see that the transition amplitude from the ground state to state $m$ is
    \begin{equation*}
        O(m) = \sum_{n=0}^{9} a(0,n) a(m,n+1) + \sum_{n=0}^{9} a(0,n+1) a(m,n)
    \end{equation*}
    Here we use the abbreviated notation $O(m) = O(0,m)$, as we will be only discussing transitions from the ground state. The wave functions $a$ are based on the values from Table \ref{tab2}. Note for each $O(m)$, the summation always starts from state 0.  We also call this as an operator T1, as what have done in the previous section and previous works \cite{1}\cite{2}\cite{3}\cite{4}\cite{5}.

    Let us first discuss transitions with the T1 operator in the asymptotic limit. The results are very simple -- all transitions vanish. This has been discussed before in refs. \cite{1}\cite{2}. One that the asymptotic Hamiltonian (\ref{2}) has no diagonal elements. Indeed we see that the transient operator T1 is proportional to the asymptotic Hamiltonian. Hence we have the commutation relationship
    \begin{equation*}
        [H_{a},\text{T1}]=0
    \end{equation*}
    This easily leads to the conclusion that there are no transitions.
    
    To see how the quantum number $s(m)$ in Eq (\ref{2.2}) come into play, we have to retreat from the extremes asymptotic condition. This is done in Table \ref{tab3} where we will go back to eigenfunctions of the full Hamiltonian (\ref{1}) and present results for $O(m)$ for various $v$'s. Note that the initial state is the lowest state (i.e. ground state) the one with $m=0$.
    
    \begin{table}[H]
        \footnotesize
        \centering
        \caption{The transition amplitude $O$ is the summation from different wave functions. $O(1)$ means the sum from state 0 to state 1, $O(2)$ means the sum from state 0 to state 2, etc. Note the values in the last row are all equal to zeros.}
        \begin{tabular}{|c||l|l|l|l|l|l|l|l|}
        \hline
        $v$ & 1,000,000 & 100,000 & 10,000 & 1,000 & 100 & 10 & 1 & 0.1 \\
        \hline\hline
        $O(1)$ & 2.1611E-6 & 2.1611E-5 & 2.1611E-4 & -2.1609E-3 & -2.1363E-2 & 1.4146E-1 & -6.1905E-1 & 9.9018E-1\\
        $O(2)$ & 1.0814E-11 & 1.0815E-9 & 1.0815E-7 & -1.0813E-5 & -1.0612E-3 & 3.9046E-2 & 7.3303E-2 & 3.2570E-4\\
        $O(3)$ & -1.7209E-7 & -1.7209E-6 & -1.7209E-5 & 1.7209E-4 & 1.7291E-3 & 1.8515E-2 & -1.3882E-2 & 4.1017E-6 \\
        $O(4)$ & 1.8371E-12 & -1.8371E-10	& -1.8371E-8 & -1.8369E-6 & -1.8134E-4 & -9.3440E-3 & -2.3145E-3 & 5.4852E-8\\
        $O(5)$ & 4.6234E-8 & -4.6234E-7 & -4.6234E-6 & -4.6236E-5 & -4.6503E-4 & 5.5790E-3 & -3.2930E-4 & 6.8679E-10\\
        $O(6)$ & -5.9153E-13 & -5.8927E-11 & -5.8926E-9 & 5.8918E-7 & -5.8211E-5 & 3.1449E-3 & -4.0919E-5 & 7.8574E-12\\
        $O(7)$ & -1.7384E-8 & 1.7384E-7 & 1.7384E-6 & -1.7385E-5 & -1.7460E-4 & 1.9424E-3 & 4.5050E-6 & 8.1909E-14\\
        $O(8)$ & -2.1104E-13 & 2.1258E-11 & -2.1259E-9 & 2.1255E-7 & -2.0954E-5 & 8.9890E-4 & -4.2558E-7 & 0\\
        $O(9)$ & 6.4267E-9 & -6.4267E-8 & 6.4267E-7 & 6.4262E-6 & 6.3750E-5 & 3.0700E-4 & -2.8344E-8 & 0\\
        $O(10)$ & 0 & 0 & 0 & 0 & 0 & 0 & 0 & 0\\
        \hline
        \end{tabular}
        \label{tab3}
    \end{table}
    
    Note that for large $v$, say from 100 up there is a simple behavior of the $O(m)$'s. If we increase $v$ by a factor of 10 e.g from 100 to 1,000. The values of $O(m)$ for odd $m$ decreases by about factor of 10 but for even $m$ they decrease by about a factor of 100. This is not true for small $v$. This only emerges asymptotically. Another striking observation from Table \ref{tab3} is the odd-even staggering of the values of $O(m)$. The odd $O(m)$'s are much larger than the even ones. One can generalize this by saying that transitions involving a change of $s(m)$ are much larger than ones that do not involve such a change.
    
    If we replace $v$ by $-v$ we get the same magnitudes for all the $O$'s, but not always the same phases. However that does not matter because the transition rate goes as $O^2$. The fact that $O(10)$ vanishes has been discussed before in \cite{1} and \cite{2}. This last point deviates from the exponential falloff and indeed has not been included in the figures in these references.
    
    \begin{table}[H]
        \footnotesize
        \centering
        \caption{Similar comparison with Table \ref{tab3} but for energy eigenvalues. In this case $E_{n}$ are not the summations but the individual values.}
        \begin{tabular}{|c||l|l|l|l|l|l|l|l|}
        \hline
        $v$ & 1,000,000 & 100,000 & 10,000 & 1,000 & 100 & 10 & 1 & 0.1 \\
        \hline\hline
        $E_0$ & -1.9319E6 & -1.9318E5 & -1.9314E4 & -1.9269E3 & -1.8842E2 & -1.6005E1 & -7.4619E-1 & -9.9506E-3\\
        $E_1$ & -1.7321E6 & -1.7320E5 & -1.7316E4 & -1.7270E3 & -1.6814E2 & -1.2289E1 & 7.8932E-1 & 1.9995E-1\\
        $E_2$ & -1.4142E6 & -1.4142E5 & -1.4137E4 & -1.4092E3 & -1.3639E2 & -8.8662 & 1.9611 & 2.0000\\
        $E_3$ & -9.9995E5 & -9.9995E4 & -9.9950E3 & -9.9500E2 & -9.4983E1 & -4.8288 & 2.9960 & 3.0000\\
        $E_4$ & -5.1763E5 & -5.1759E4 & -5.1714E3 & -5.1264E2 & -4.6756E1 & -1.0104E-1 & 3.9998 & 4.0000\\
        $E_5$ & 5.0000 & 5.0000 & 5.0000 & 5.0000 & 5.0000 & 5.0000 & 5.0000 & 5.0000\\
        $E_6$ & 5.1764E5 & 5.1769E4 & 5.1814E3 & 5.2264E2 & 5.6756E1 & 1.0101E1 & 6.0002 & 6.0000\\
        $E_7$ & 1.0000E6 & 1.0000E5 & 1.0005E4 & 1.0050E3 & 1.0498E2 & 1.4829E1 & 7.0040 & 7.0000\\
        $E_8$ & 1.4142E6 & 1.4143E5 & 1.4147E4 & 1.4192E3 & 1.4639E2 & 1.8866E1 & 8.0389 & 8.0000\\
        $E_9$ & 1.7321E6 & 1.7321E5 & 1.7326E4 & 1.7370E3 & 1.7814E2 & 2.2289E1 & 9.2107 & 9.0000\\
        $E_{10}$ & 1.9319E6 & 1.9319E5 & 1.9324E4 & 1.9369E3 & 1.9842E2 & 2.6005E1 & 1.0746E1 & 1.0001E1\\
        \hline
        \end{tabular}
        \label{tab4}
    \end{table}
    
    As shown in Table \ref{tab4} the energies are approximately proportional to $v$ for very large $v$ (this is not the case for small $v$).

\section{Perturbation theory in the strong coupling limit}

    In previous works \cite{3}\cite{4}\cite{5} we considered perturbation theory in the weak coupling limit. That meant that the $v$'s and $w$'s were small compared to $nE$. For example, in the tridiagonal case we found for the ground state that
    \begin{equation*}
         a(0,n) = \frac{(-1)^{n} v'^{n}}{n!}
    \end{equation*}
    where $v'=v/E$. Selected results for the pentadiagonal case \cite{5} are: $a(0,0)=1$, $a(0,1)= -v'$, $a(0,2)=v'^{2}/2-w'/2$, $a(0,3) =-v'^{2}/6+v'w'/2$. We were able to get analytic expressions for the $a(m,n)$'s in terms of the $v$'s and $w$'s. We were able to go further and display analytically the exponential behaviors seen in the figures. 
    
    We now wish to consider perturbation theory in the strong coupling limit. We will limit ourselves to the tridiagonal case. Our motivation for doing this is to explain the striking results in the previous section that are seen as we increasingly vary $v$. From Table \ref{tab4} we have noted that the energies for all states increase linearly with $v$. From Table \ref{tab3} we see that for transitions from the ground state to odd $n$ states the T1 transition amplitudes $O_n$ are linear in $v$ whilst transitions from ground to even $n$ states appear to be quadratic in $v$. We will now demonstrate that those results can be explained by using what we call strong coupling perturbation theory.

    To this end we turn things around and say that the diagonal terms $nE$ are small compared to the $v$'s. We further utilize the fact that for T1 all transitions for the Hamiltonian $H_{a}$ (the one for which the diagonal terms are set to zero) vanish. We note that although the matrix elements of $H_a$ depend on $v$, the eigenfunctions do not. The reason for this is that $H_a$ can be written as $vH'_a$ and the latter matrix does not depend on $v$. Rather $H'_a$ consist of ``ones'' on the two off-diagonals. The eigenfunctions of $H_a$ are the same as those of $H'_a$ while the eigenvalues of $H_a$ are $v$ times those of $H'_a$. This explains the results of Table \ref{tab4}. In that table we have not reached the extreme limit of $H_{a}$ but as we approach the results become increasingly qualitatively the same.

    For the first order perturbation theory,
    \begin{equation}
       \vert f\rangle \cong \vert f^{0}\rangle + \sum \frac{\langle s^{0}\vert V_{pert}\vert f^{0}\rangle}{E_{f}^{0} - E_{s}^{0}} \vert s^{0}\rangle
    \label{4.1}
    \end{equation}
    , where $|f^{0}\rangle$ are the eigenstates of $H_a$ of Eq (\ref{2}). For conventions, here $f$ means \textit{fixed} and $s$ means \textit{summed} index. Also, the $V_{pert}$ here is the matrix with $nE$ along the diagonal and nothing else. Note that the numerator $\langle s^{0}|V_{pert}|f^{0} \rangle$ does not depend on $v$. All the $v$ dependence is in the energy denominator.

    In Table \ref{tab4} we show the eigenvalues of $H_c$ Eq (\ref{1}). Of course for $v=0$ they will be the diagonal terms $E_{n}=nE$ (with $E$ taken to be one). The behavior for small $v$ is somewhat complicated but for large $v$ they are almost exactly proportional to $v$. Thus the energy denominators in Eq (\ref{4.1}) are also proportional to $v$ for large $v$.

    Now we are going to consider the energy denominators in Eq (\ref{4.1}). Setting $E = 1$ for convenience, so that $nE = n$. We first do the states 0, 1, 2:
    \begin{equation}
    \begin{split}
         \langle 0\vert V_{pert} \vert 1\rangle &= \sum_{n}^{10} a(0, n) a(1, n) nE 
        \\
        \langle 0\vert V_{pert} \vert 2\rangle &= \sum_{n}^{10} a(0, n) a(2, n) nE 
    \end{split}
    \end{equation}
    Here $V_{pert}$ means the perturbative coupling parameter. We employ the results from Table \ref{tab1} and then obtain
    \begin{equation}
    \begin{split}
        \langle 0\vert V_{pert} \vert 1\rangle &= -2.16113\\
        \langle 0\vert V_{pert} \vert 2\rangle &= 0
    \end{split}
    \end{equation}
   More generally we have the striking fact that $\langle m\vert V_{pert} \vert m'\rangle$ vanishes for all cases where the quantum number $s(m)$ is the same as $s(m')$. Indeed the non-vanishing value of $\langle n\vert V_{pert} \vert m\rangle$ are those for which $n$ and $m$ have different quantum numbers. This can explain why the $O(\text{0 to }m)$'s in Table \ref{tab3} are proportional to $1/v$ for odd $m$ and $1/v^2$ for for even $m$. The energy denominators in Eq (\ref{4.1}) are approximately proportional to $v$ for large $v$. This explains the $1/v$ behavior of $O(m)$ for odd $m$. But since $\langle 0\vert V_{pert} \vert m\rangle$ vanishes for even $m$ we have to go beyond first order perturbation theory to get any effect. In second order perturbation theory we have two energy denominators and so we pick up next's factor of $1/v$ -- hence $1/v^2$. Also since we get no contribution to $O$ in zero and first order perturbation theory, the magnitude of $O(0,m)$, i.e. a transition from the ground state to the $m$'th excited state is much smaller for even $m$ than for odd $m$ -- hence even-odd staggering. Note that for positive $v$ the coupling term $\langle 0\vert V_{pert} \vert m\rangle$ is independent of $v$. This is because the eigenfunctions therein are from the Hamiltonian $H_a$.

    Armed with the perturbation expression Eq (\ref{4.1}), we can then construct the new wave functions, denoted by $d(N,n)$ and satisfies
     \begin{equation}
        \vert m\rangle = \sum_{n=0}^{10} d(m, n) \vert n\rangle
    \end{equation}
    Note the $m$ values are fixed from 0 to 10. Alternatively, for individual $d(m,n)$, we can simply express as:
   \begin{equation}
       d(m,n) = a(m,n) + \frac{\langle m \vert V_{pert}\vert n \rangle}{E_{m} - E_{n}}, \quad n \text{= 0, 1, ..., 10}
    \label{4.5}
    \end{equation}
    The results we get for the new wave functions derived from perturbation theory are shown as Table \ref{tab5}. Compared with Table \ref{tab2}, we find that both wave functions are pretty similar and almost like the same. It suggests that our perturbation method can give sufficiently accurate results.

    \begin{table}[H]
    \centering
    \caption{The new wave functions $d(m,n)$ for $v=10^4$, $w=0$, constructed by the perturbation calculations Eq (\ref{4.5}).}
    \includegraphics[width=1\textwidth]{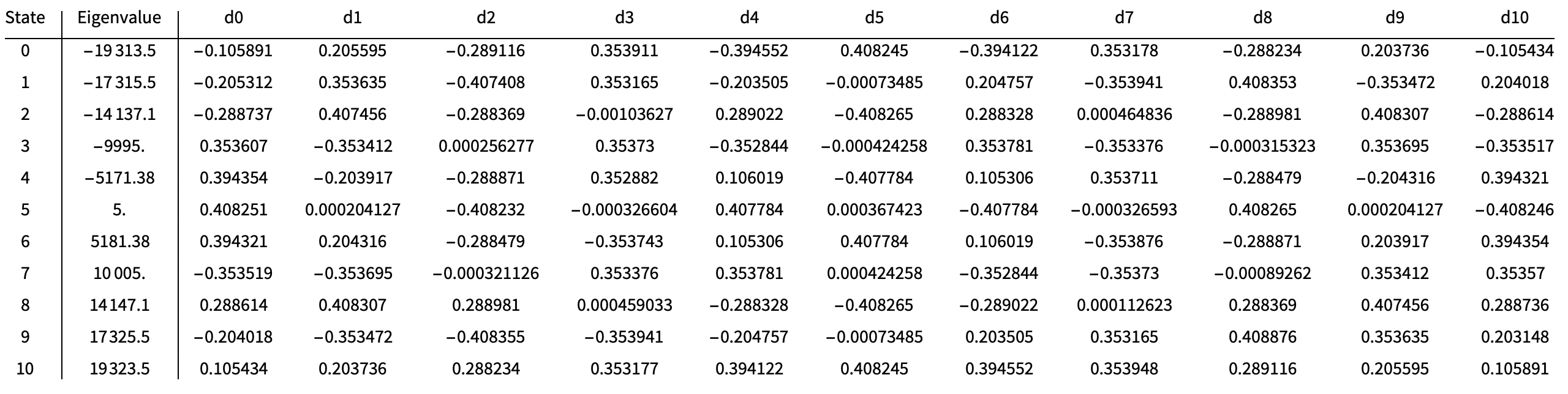}
    \label{tab5}
    \end{table}

    In the weak coupling model the quantum number for the state is $n$ and the energy is $nE$. In our strong coupling limit the new quantum number is given in Eq (\ref{2.2}), $s(m) = a(m,10)/a(m,0)$. There are other examples where there are changes in quantum numbers as we go from weak coupling to strong coupling. For example, there is the Zeeman effect for atomic hydrogen. In the $1p$ shell in the limit of a weak magnetic field the good quantum numbers are $J=m_{J}$. The $1p_{\frac{3}{2}}$ shell and $1p_{\frac{1}{2}}$ shell are well separated. In the strong coupling limit, i.e. large magnetic field, the good quantum numbers are $m_l$ and $m_s$. Most states consist of admixtures of both $J=\frac{1}{2}$ and $J=\frac{3}{2}$.

\section{Closing remarks}
    We have shown in this and previous works \cite{1}\cite{2}\cite{3}\cite{4}\cite{5} we can get very interesting outcomes from rather simple matrices-both in the weak coupling and the strong coupling limits. In this paper, we demonstrated the symmetric patterns among the values and introduced a quantum number to classify the odd-even results under strong coupling. Moreover, we showed that for large $v$, a transition from the ground state to an even $m$ excited state was much larger than to a neighboring even $m$ state. Meanwhile, showed that the effects would disappear when decreasing the coupling parameter from $100$ to $10$ and below. Eventually, we performed the first-order perturbation theory and confirmed that our results are sufficiently consistent. In the near future, we hope to consider other examples, e.g. more work with pentadiagonal matrices in the strong coupling limit. We hope others will follow our example. Several of the works cited here can also be found in the archives under Zamick: arXiv:2009.04463, arXiv:1907.11528, arXiv:1902.03119, arXiv:1811.02562, arXiv:1807.08552, and arXiv:1803.00645.

\section{Acknowledgment}
C. Fan acknowledges the supports from Aresty Research Center. We thank Mingyang Ma for her help.

\end{document}